# Precise determination of the spectroscopic $g$–factor using broadband ferromagnetic resonance spectroscopy*

Justin M. Shaw[1], Hans T. Nembach[1,2], T.J. Silva[1], and Carl T. Boone[1]

[1]Electromagnetics Division, National Institute of Standards and Technology, Boulder, Colorado 80305

[2]JILA, University of Colorado and NIST, Boulder, Colorado 80309, USA

We demonstrate that the spectroscopic $g$–factor can be determined with high precision and accuracy by broadband ferromagnetic resonance measurements and applying an asymptotic analysis to the data. Spectroscopic data used to determine the $g$–factor is always obtained over a finite range of frequencies, which can result in significant errors in the fitted values of the spectroscopic $g$–factor. We show that by applying an asymptotic analysis to broadband datasets, precise values of the intrinsic $g$–factor can be determined with errors well below 1 %, even when the exact form of the Kittel equation (which describes the relationship between the frequency and resonance field) is unknown. We demonstrate this methodology with measured data obtained for sputtered $Ni_{80}Fe_{20}$ ("Permalloy") thin films of varied thicknesses, where we determine the bulk $g$–factor value to be 2.109 ± 0.003. Such an approach is further validated by application to simulated data that includes both noise and an anisotropy that is not included in the Kittel equation that was used in the analysis. Finally, we show a correlation of thickness and interface structure to the magnitude of the asymptotic behavior, which provide insight into additional mechanisms that may induce deviations from the Kittel equation.

**Introduction**

The response of magnetic materials to microwave excitations is an ongoing subject of intense technological concern, given the inherent ability of ferromagnetic spins to react to perturbations at microwave frequencies and beyond. Most recently, a large number of studies of magnetic damping in ferromagnetic metallic alloys have been driven by interest in developing non-volatile, high-speed, low-power magnetic random access memories (MRAM) that utilize spin-transfer torque (STT) to set the bit state.[1,2] Such memories show promise to eventually replace DRAM and even SRAM in microprocessors.[3] The efficiency of the write process for STT-MRAM is constrained by the magnetic damping parameter, which dictates the rate at which angular momentum can be exchanged between the spin system and the crystal lattice. Thus, the discovery of magnetic memory materials with the lowest possible damping has become a prime concern of STT-MRAM developers.

Of equal importance is the ability to engineer high anisotropy materials to increase the thermal stability of STT-MRAM. Magnetic anisotropy energy (MAE) is fundamentally related to the spin-orbit coupling (SOC) of the material that results in an energetic dependence between the spin system and material structural asymmetries that results in a favored axial orientation of the spins. Thus, there is additional interest in measurement techniques that can characterize the strength of the SOC and how it is correlated to measured values of MAE.[4–8] One means of determining the SOC in ferromagnets is to measure the contribution of orbital magnetization to the total moment. In the absence of SOC, the orbital moment is quenched for any crystal structure with cubic symmetry. SOC overcomes such quenching that results in an orbital moment that scales in proportion to the ratio of the SOC and the interatomic ligand field.[9] Thus, measurement of the orbital moment is a direct means of determining the SOC.

Ferromagnetic resonance spectroscopy (FMR) is a powerful measurement technique for characterizing magnetic materials in virtue of its ability to simultaneously measure the Landau-Lifshitz damping parameter, the spectroscopic $g$–factor, anisotropy, and magnetic inhomogeneity.[10] Traditionally, FMR is performed by loading magnetic samples into single-frequency microwave cavities.[11] More recently, broadband FMR techniques have become increasingly popular and accessible. In this case, the microwave excitations are delivered to the sample by either a strip-line or coplanar waveguide (CPW) structure. Broadband microwave generation and detection can be performed by use of either a microwave signal generator in conjunction with a diode detector, or with a vector network analyzer that acts as both the source and phase-sensitive detector of the requisite microwaves.[12] In addition, time domain measurements can also be used as an alternative to FMR in the frequency domain. Such time domain measurements can be performed by use of pulsed-inductive microwave magnetometry (PIMM),[12-15] optical pump-probe technique,[16-20] [refs] or a synchronized pulsed laser technique.[21] Although not a direct measurement of the dynamic susceptibility, the relationship between the resonance field and frequency can also be determined by use of Brillouin light scattering (BLS).[22-24]

All of these discussed techniques can, in principle, determine the spectroscopic $g$–factor by extraction of the proportionality of the gyromagnetic frequency to the net internal field $H_i$ acting on the spins, given by $\omega = (\gamma \mu_0 H_i)$, where $\gamma = (g \mu_B / \hbar)$, $\mu_0$ is the permeability of free space, $\mu_B$ is the Bohr magneton, $\hbar$ is the reduced Planck's constant, and $g$ is the spectroscopic $g$–factor. Determination of $g$ is valuable since the relative spin and orbital moments of a material can be evaluated by use of the well-known relation,[9]

$$\frac{\mu_L}{\mu_s} = \frac{g-2}{2},$$

(1)

where, $\mu_L$ is the orbital moment *per spin* and $\mu_S = \mu_B$ is the spin moment.

This capability is significant since x-ray magnetic circular dichroism (XMCD) at synchrotron facilities are more commonly used to evaluate the spin and orbital moments of materials. Thus, for a large variety of samples and experiments, an FMR approach can also be used. As an example, the relationship between the orbital moment asymmetry and the perpendicular anisotropy was determined from FMR measurements in Fe/V[5] and $Co_{90}Fe_{10}$/Ni[8] mulitlayers. In these studies, the origin of the magnetocrystalline anisotropy was related at a fundamental level to the asymmetry in the orbital moment, confirming theoretical predictions.[7,25]

While this relationship has been well established for many decades, the evaluation of the orbital moment via FMR has been largely inhibited due to the difficulty in determining the $g$–factor with less than 1 % error and/or requiring dependent assumptions about the anisotropy or saturation magnetization of the sample.[10] Whereas a 1 % error in the total $g$–factor appears as a relatively small error, such an error can translate into significant uncertainty in the measurement of the orbital or spin moments of a material; easily obscuring changes in the orbital moment that result from variations in the electronic structure and anisotropy.[26] As an example, reported values of the $g$–factor for Permalloy (taken as either $Ni_{80}Fe_{20}$ or $Ni_{81}Fe_{19}$) range from 2.0 to 2.17; exceeding an 8 % variation among studies .[9,27-32] Since the $g$–factor is an intrinsic material property, such a large variation may indicate that the measurement parameters may strongly influence the determined value of $g$. In fact, in one study, the fitted value of the $g$–factor changed from 2.08 when measured at 10 GHz to 2.12 when the same sample was measured at 24 GHz.[9] This report presents one of the first indications that different values of the $g$–factor may be measured depending solely on the measurement parameters.

In this paper, we explicitly show that different values of the $g$–factor can be obtained due to a dependence on the range of the applied magnetic field; possibly explaining why there is variation among reports. More importantly, we report on a new asymptotic analysis method for determining a precise value of the $g$–factor that overcomes the limitations imposed by finite frequencies and fields available in the laboratory. This method makes use of broadband FMR combined with an asymptotic analysis of the data as the frequency $f \to \infty$ to increase the *precision*. By taking into account the systematic errors in the field and frequency calibrations, an accurate value of the $g$–factor is determined.

**Experiment**

Thin film $Ni_{80}Fe_{20}$ ("Permalloy") samples of varying thickness were dc magnetron sputter deposited directly from a $Ni_{80}Fe_{20}$ target with an Ar pressure of approximately 0.5 mTorr (0.07 Pa). Prior to the $Ni_{80}Fe_{20}$ deposition, a 3 nm Ta seed layer was deposited to ensure good adhesion and a high-quality {111} texture. Samples were capped with a sputter-deposited 5 nm $Si_3N_4$ layer to prevent oxidation. Samples were rotated during growth to minimize the in-plane anisotropy, which was verified to be less than 0.3 mT using magnetometry.

The utilized FMR spectrometer employs a room-temperature-bore superconducting magnet capable of applying fields as large as $\mu_0 H$ = 3 Tesla. We use a broadband (1—70 GHz) vector network analyzer to apply microwave fields to the sample via a CPW. The samples are first coated with a thin polymethyl methacrylate (PMMA) insulating layer to avoid shorting the CPW when placed face down for measurements. The transmission parameter $S_{21}$ is then measured at fixed frequency as a function of applied magnetic field that is ramped from the highest field value to the lowest field value. (Details of the measurement technique are provided in Ref. [33]. ) Figure 1 contains examples of measured complex spectra as the magnetic field is swept through the resonance for 8 GHz and 55 GHz. These resonances are described by the dynamic susceptibility derived from the Landau-Lifshitz equation, which is used to fit the data as outlined in Ref [33]. From these fits, we obtain values of the resonance field $H_{res}$ and field-swept linewidth $\Delta H$ for each frequency. The relationship between frequency and $H_{res}$ is described by the Kittel equation, which is for the in-plane field geometry is,[8,34]

$$f(H_{res}) = \frac{g\,\mu_0\mu_B}{h}\sqrt{(H_{res} + H_k^a(\varphi))(H_{res} + M_{eff} + H_k^b(\varphi))} \qquad (2)$$

where $h$ is Planck's constant, $H_k^a$ and $H_k^b$ are the in-plane anisotropy fields, which can contain terms of multiple symmetries, $\varphi$ is the azimuthal angle, and the effective magnetization $M_{eff}$ is given by,

$$M_{eff} = M_s - \frac{2K}{\mu_0 M_s} \qquad (3)$$

where $M_s$ is the saturation magnetization and $K$ is the perpendicular (out-of-plane) anisotropy energy density. We use a sign convention whereby a positive value of $K$ favors a perpendicular magnetization. Data for the frequency dependence of $H_{res}$ in the case of a 10 nm thick sample, as well as the fit to Eq. (2) are presented in Fig. 1.

**Results**

*Experimental in-plane data*

Examination of Eq. (2) specifies that the $g$–factor can be determined by a simple fit of the FMR data regardless of the range. This is most easily observed when $H_{res} \gg M_{eff}$ and $H_{res} \gg H_k^{a,b}$ since $g$ is simply the proportionality constant between $f$ and $H_{res}$ (assuming a fully saturated sample). However, since $\mu_0 M_{eff}$ for $Ni_{80}Fe_{20}$ is $\approx 1$ T, the frequencies needed before this condition can be met will be in excess of 200 GHz. FMR measurements at such frequencies are outside a reasonably obtainable capability in the laboratory. At lower fields and frequencies, $f$ and $H_{res}$ are no longer proportional and the accuracy of the $g$–factor is highly dependent on the values determined for $M_{eff}$ and $H_k^{a,b}$. Other factors such as the degree to which the sample is fully saturated will also affect the accuracy of the $g$–factor at lower fields. Since every FMR spectrometer has different capabilities depending on the frequencies and fields available to the instrument, these later points may introduce variations in the fitted values of $g$.

Given the nonlinear dependence of $f$ on $H_{res}$ in Eq. (2), both the precision and accuracy of the extracted values for $M_{eff}$, $H_k^{a,b}$, and $g$ from fitting of data to Eq. (2) are necessarily dependent on details of both the dc permeability (i.e. how much field is required to fully saturate the magnetization) and the frequency range of the fit. For example, if the fitting range is in the limit of $H_k \ll H_{res} \ll M_{eff}$, then Eq. (2) reduces to $f(H_{res}) \sim [(g\mu_0 \mu_B)/h] (H_{res} M_{eff})^{1/2}$ and it is no longer possible to extract $g$ and $M_{eff}$ independently from each other. As such, we should expect that the errors in the fits should decrease rapidly with increasing fitting range. We therefore first examine how the finite fitting-range influences both the error and the value for $g$ obtained from a nonlinear least-squares fit of the data to Eq. (2).

The resonance field for the 50 nm $Ni_{80}Fe_{20}$ sample was measured from 4 GHz to 60 GHz in 1 GHz increments for the in-plane geometry. We first define two frequencies: the lower fitting frequency $f_{low}$ (or lower bound on the data used in the fit), and the upper fitting frequency $f_{up}$ (or upper bound to the data used in the fit). The dataset for the 50 nm $Ni_{80}Fe_{20}$ sample is then fit to Eq. (2) with $f_{low}$ = 4 GHz for all of the fits. Since our samples have negligible in-plane anisotropy, we have set $H_k^a = H_k^b = 0$ in Eq. (2) for these fits. We vary $f_{up}$, starting at 20 GHz, and increasing with 1 GHz increments up to 60 GHz. The fitted values of $g$ (which we define as $g_{fit}$ to distinguish from the intrinsic value of $g$) for each value of $f_{up}$ are plotted in Fig. 2. (For clarity, the insets of Fig. 2 demonstrate the data that were fit to determine $g_{fit}$ for a few points on the curve. The values of $f_{low}$ and $f_{up}$ are indicated in the inset plots.) While we see that the estimated error bars for the fits decrease with increasing $f_{up}$, as expected, *we also find that $g_{fit}$ varies substantially over a range of 2.02 to 2.10*, with a strong dependence on the value of $f_{up}$. Indeed, we also find that $g_{fit}$ approaches an asymptotic value as $f_{up}$ is increased, with the implication that application of a large enough field/frequency leads to a value for $g_{fit}$ that is approximately independent of frequency range. As we will see, this later point forms the basis of our analysis and our assumption that the most precise determination of $g$ is the asymptotic value as $f_{up} \to \infty$, as indicated by the horizontal line in Fig. 2.

The next question we address concerns the functional form for the dependence of the extracted value of $g_{fit}$ on $f_{up}$. We have empirically found that the data for $g_{fit}$ in Fig. 2 is approximately a linear function of $1/f_{up}^2$. Such a power law dependence can be verified in Fig. 3(a) where we plot $\log(g_{fit} - g)$ versus $\log(f_{up})$, where, $g$ is the asymptotic value of the $g$–factor. The slope of this line was consistently found to be $-2.0 \pm 0.2$ among samples. We present a plot of the fitted value for $g_{fit}$ as a function of $1/f_{up}^2$ in Fig. 3(b). There is only a slight deviation from linearity for $f_{up} \leq$ 23 GHz. We determine the asymptotic value of $g$ from the $y$-intercept (also indicated as the horizontal line in Fig. 2) via a linear regression fit to the data in Fig. 3(a). In addition, we find that $M_{eff}$ has a similar linear dependence on $1/f_{up}^2$, as presented in Fig. 3(c).

For the analyses presented so far, $f_{low}$ has been held fixed at 4 GHz. In Fig. 4(d), we present a series of plots of $g_{fit}$ as a function of $1/f_{up}^2$, with three different values of $f_{low}$: 4 GHz, 8 GHz, and 12 GHz. (We have omitted error bars in the figure for clarity.) The $y$-intercepts for the three curves are $2.111 \pm 0.002$, $2.109 \pm 0.001$, and $2.108 \pm 0.001$, respectively. Thus, we see that our asymptotic analysis method is relatively insensitive to the exact value of $f_{low}$. Fig. 4(d) also shows that as $f_{low}$ is increased, the magnitude of the slope in the $g$ vs. $1/f_{up}^2$ plot decreases. (The linear regression fits are weighted to the error bars. This becomes important since there is some deviation from linearity at lower frequency values.) This behavior indicates that the mechanism responsible for large variation in fitted $g$ is primarily confined to the lower frequencies. By taking the average of $g$ (weighted to the respective error bars for each fit) determined from several values of $f_{low}$ (4 GHz, 8 GHz, 12 GHz, and 16 GHz), we determine a value of $2.108 \pm 0.002$ for the 50 nm film, where the error is the standard deviation among these values.

We also performed identical measurements and analyses with $Ni_{80}Fe_{20}$ films of varying thickness, with the results shown in Figs. 4(a), 4(b), and 4(c) for film thicknesses of 5 nm, 10 nm, and 20 nm, respectively. The same asymptotic behavior is observed in each of the other samples, with a negligible dependence of the *y*-intercept value on $f_{low}$. However, the slope of $g_{fit}$ vs. $1/f_{up}^2$ depends on sample thickness. The 20 nm $Ni_{80}Fe_{20}$ sample data has a smaller slope than that for the 50 nm $Ni_{80}Fe_{20}$ sample, and the 10 nm $Ni_{80}Fe_{20}$ is yet smaller. The slope even changes sign for the 5 nm $Ni_{80}Fe_{20}$ data.

The values of *g* and $M_{eff}$ determined by this asymptotic method for all four $Ni_{80}Fe_{20}$ samples are plotted as a function of the reciprocal film thickness $1/t$ in Figs. 5(a) and 5(b), respectively. We see a small $1/t$ dependence of *g* and $M_{eff}$, as to be expected in the presence of an interfacial anisotropy.[5,8] From the *y*-intercept of the $1/t$ dependence, we obtain the extrapolated *bulk* values for $Ni_{80}Fe_{20}$ of $g = 2.109 \pm 0.001$ and $\mu_0 M_{eff} = 1.019 \pm 0.002$ T, where the values for the uncertainty only represent the precision of the measurement. By taking into account the additional uncertainty in the magnetic field calibration, we determine with accuracy the *bulk* values for $Ni_{80}Fe_{20}$ to be $g = 2.109 \pm 0.003$ and $\mu_0 M_{eff} = 1.019 \pm 0.003$ T. We compare this value of *g* to those previously reported for Permalloy in Table I. To aid in the comparison, we also include in Table I the thickness and method used to determine *g*. The value of $2.109 \pm 0.003$ that we measure is within the error bars of those previously reported for 80 % of the references listed in Table I.

**Table I.** Previously reported values of the *g*–factor for Permalloy along with the method and frequency range used for the measurement. For magnetomechanical measurements that utilized the Einstein-de Haas effect, *g* was determined from measurements of the magnetomechanical factor $g'$ via the relation $1/g + 1/g' = 1$.

| Permalloy Thickness | *g*–factor | Method | Frequency range | Reference |
|---|---|---|---|---|
| **5 –50 nm** | 2.109 ± 0.003 | VNA-FMR with | 4—60 GHz | This work |

| (extrapolated to bulk value) | | asymptotic analysis | | |
|---|---|---|---|---|
| 50 nm | 2.20 ± 0.12 | Einstein-de-Haas | — | [27] |
| 50 nm | 2.0—2.1 | PIMM | 0—2 GHz | [28] |
| 50 nm | 2.1 | PIMM | 0—3 GHz | [29] |
| 10 nm | 2.05 | PIMM | 0—3 GHz | [29] |
| bulk | 2.12 ± 0.02 | FMR | 19.5 & 26 GHz | [30] |
| bulk | 2.12 | Einstein-de-Haas | — | [31] |
| 4—50 nm | 2.08 ± 0.01 | FMR | not reported | [32] |
| bulk | 2.08 ± 0.03 | FMR | 10 GHz | [9] |
| bulk | 2.12 ± 0.03 | FMR | 24 GHz | [9] |

*Inclusion of an in-plane anisotropy term*

In the analysis presented so far, we have neglected the in-plane uniaxial anisotropy since magnetometry indicated that it is less than 0.3 mT in magnitude. We now include a uniaxial anisotropy $H_k$ term in Eq. (2) as a free parameter in the fit, where we take $H_k = H_k^a = H_k^b$. Figure 6(a) is a plot of the fitted value of $g_{fit}$ versus $f_{up}$ for the 50 nm $Ni_{80}Fe_{20}$ film. The variation of $g_{fit}$ is reduced relative to the fits that neglect $H_k$, but still exhibits an asymptotic behavior. By use of a plot of $\log(g_{fit} - g)$ vs. $\log(f_{up})$ [see inset in Fig. 6(b)], a value of −1.2 is obtained for the exponent in the power law that describes the asymptotic approach in contrast to the previously determined value of approximately −2.0 when $H_k$ is omitted from Eq.(2). A plot of $g_{fit}$ versus $1/f_{up}^{1.2}$ is presented in Fig. 6(b). The asymptotic approach is complicated by the strong presence of undulations in the data and therefore does not exhibit a smooth trend. Nevertheless, a linear regression fit to the data in Fig. 6(b) yields a value of $g = 2.110 \pm 0.004$, which is within error of the previously determined value for the 50 nm $Ni_{80}Fe_{20}$ film. However, the presence of the undulations prevents an analysis that takes into account multiple value of $f_{low}$, since when the fitting range is reduced, the undulations obscure the trend of the data.

Comparison of the results obtained from fits that either include or exclude $H_k$ in the Kittel equation suggest that a more accurate values of $g$ can be obtained by employing an independently estimated, non-zero value for $H_k$ in the analysis, even when there is significant uncertainty in the value for $H_k$. Inclusion of $H_k$ as an additional fitting parameter in Eq. (2) results in three coupled adjustable parameters that are not orthogonal to each other during the least-squares non-linear fitting process. Recall that the purpose of our analysis in the first place was to disentangle the 2 coupled fitting parameters $M_{eff}$ and $g$. Thus, this process is more susceptible to systematic errors by the inclusion of an additional fitting parameter, despite the decrease in the residue between the data and the *individual* fits to Eq. (2) (as would be expected when increasing the number of fitting parameters).

Figure 6(c) shows a plot of the fitted values of $H_k$ as a function of $f_{up}$ as well as the error in $H_k$ obtained from the fit. The fitted values of $\mu_0 H_k$ vary from 1.2 to 1.4 mT over this range with large error bars relative to the magnitude. As a point of comparison, we alternatively determine $H_k$ by fitting the data with both $g$ and $M_{eff}$ fixed to their asymptotic values. (i.e. $H_k$ becomes the only fitting parameter). In this case, the value of $\mu_0 H_k$ determined from the fit becomes 0.2 ± 0.1 mT, which is consistent with the value of < 0.3 mT obtained from the magnetization curves shown in Fig. 6(d). This suggests that the best protocol for determining all three parameter ($M_{eff}$, $g$, and $H_k$) is to first apply the asymptotic analysis to determine $M_{eff}$ and $g$ with $H_k$ set to a fixed value, then perform a second fit to determine $H_k$ with $M_{eff}$ and $g$ fixed to their asymptotic values.

*Measurements in the out-of-plane geometry*

So far, we have restricted the discussion to in-plane measurements of the $g$–factor. However, we now show that this analysis is equally applicable to measurements in the out-of-plane geometry. The Kittel equation for the out-of-plane geometry is,

$$f(H_{res}) = \frac{g^{\perp} \mu_0 \mu_B}{h}\left(H_{res} - M_{eff}^{\perp}\right),\tag{4}$$

where $g^\perp$ and $M_{eff}^\perp$ are used to distinguish the *g*–factor and effective magnetization measured in the out-of-plane direction, respectively. Of significance is the fact that the in-plane anisotropy fields are absent in the Kittel equation for this geometry. Figure 7 shows plots of the fitted out-of-plane *g*–factor as a function of $1/f_{up}^2$ for the 10 nm and 20 nm thick samples, which also exhibit the same asymptotic behavior observed for the in-plane data. The asymptotic values of $g^\perp$ and $M_{eff}$ are included in Fig. 5. The values of $M_{eff}$ for the in-plane and out-of-plane measurements are *identical* for all thicknesses, indicative that our methodology is self-consistent for in-plane and out-of-plane measurements. In contrast, $g^\perp$ is smaller than the in-plane *g*–factor for all thicknesses. In addition, the dependence of $g^\perp$ on reciprocal thickness is opposite in sign. However, with the presence of a perpendicular interface anisotropy (indicated by the linear dependence of $M_{eff}$ on reciprocal thickness), an asymmetry of the orbital moment is predicted and expected to increase in magnitude as the thickness of the magnetic layer decreases.[5,8,25] A linear fit to the data yields a bulk asymptotic value of 2.111 ± 0.003 for the *y*-intercept, which is within error of the value obtained for the in-plane geometry (*g* = 2.109 ± 0.001). We emphasize that such a thickness- and geometry-dependence of the spectroscopic *g*–factor may have contributed to variations of values for *g* reported in the literature, given that such factors are generally not taken into consideration, especially for thin films.

*Analysis of Simulated Data*

Without proof, we have made the assumption that the intrinsic value of *g* for the material is the asymptotic value obtained from our analysis methods. To further validate this assumption, we used simulated data with predetermined values of *g*, $M_{eff}$, and $H_k$, to which we then applied the same analysis approach. We added random white noise to the data and then used Monte Carlo-like methods to analyze the results. All simulated data are generated by use of Eq. (2) with *g* = 2.11 and $\mu_0 M_{eff}$ = 1 T. In order to add noise to the resonance field, we define a noise amplitude $N_H$ and add a randomly generated field value that lies between ± $N_H$ for each value of the resonance field.

We first apply the asymptotic analysis to the case of $\mu_0 H_k$ = 3 mT, but neglect $H_k$ in the Kittel equation used in the fitting routine (i.e. set $H_k$ = 0 for the fits). Figure 8(a) shows plots of the fitted $g$ versus 1 / $f_{up}$ before noise was added to the data ($N_H$ = 0). As observed with the experimental data, these data are linear, have slopes that decrease as $f_{low}$ is increased, and all the curves have similar y-intercepts. Thus, the asymptotic behavior we observed in the experimental data can be largely reproduced by negelecting a magnetic anisotropy in the Kittel equation that is present in the sample. From this analysis, a value of 2.110± 0.002 is determined for $g$, despite the fact that $H_k$ was neglected in the fitting Kittel equation.

We repeat the same analysis by adding noise of amplitude $N_H$ = 2 mT to the data in Fig. 8(b). The presence of the noise is revealed by the increased scatter of the data. In this case, a value of 2.113 ± 0.002 is determined for $g$, which is within 0.14 % of the value of 2.11 used to generate the simulated data. Again these data indicate that a very accurate value of $g$ can be obtained when the Kittel equation does not include all the anisotropy fields.

To demonstrate the robustness of this analysis approach, we repeat the analysis with several dataset where both $N_H$ and $H_k$ are varied. Figures 9 show a plot of the values of $g$ obtained from the asymptotic analysis for the case when $H_k$ is excluded as a fitting parameter. The data are scattered about the value of $g$ = 2.11 within approximately ± 0.1 %.

Next we demonstrate the case where $H_k$ is included as a free parameter in the fitting procedure. The analysis for this case is shown in Fig. 8(c) for $N_H$ = 2 mT . The presence of the noise in the data causes a large variation in $g_{fit}$. More importantly, these variations take the form of undulations, similar to that observed in the experimental data when $H_k$ was included as a fitting parameter [Fig. 6(b)]. These undulations dominate the behavior and do not exhibit a clear or consistent trend in the dependence on $f_{up}$. As a result, the asymptotic analysis cannot be performed in this case.

**Discussion**

We have shown that the asymptotic behavior results from systematic errors that are incurred when the actual stiffness fields present in the physical system under investigation, are omitted from the Kittel equation used to fit the data. This was most clearly demonstrated by the exclusion of the in-plane uniaxial anisotropy term in the Kittel equation for both the experimental and simulated data. However, even when in-plane anisotropy is included as a fitting parameter, we still observe a non-trivial asymptotic trend of the fitted value for $g$ as $f \to \infty$, albeit with a lessened dependence on $f_{up}$. In addition, a similar asymptotic trend is observed for the out-of-plane geometry, where in-plane, uniaxial anisotropy should have a negligible effect. One possible explanation is that the Kittel equation with only uniaxial- and easy-plane-anisotropy fields is inadequate as a complete description of the functional dependence between $f$ and $H_{res}$. There are several physical mechanisms that may give rise to such inadequacies: (1) small misalignment of the sample plane with respect to the externally applied magnetic field that results in field-dragging;[35,36] (2) a large saturation field on the order of the applied field due to spin pinning at defects or interfaces; and (3) additional anisotropies in the $Ni_{80}Fe_{20}$ samples such as higher order anisotropies or a rotatable anisotropy.

One clue as to the mechanism that governs this behavior lies in the fact that the slope (for a given value of $f_{low}$) of the data shown in Fig. 4 varies monotonically with the thickness of the $Ni_{80}Fe_{20}$ layer. This indicates that there is a thickness-dependence of material properties that gives rise to this phenomenon. In addition, for a given sample, the slope of the data in Fig. 4 consistently decreases as $f_{low}$ is increased, indicating that the effect is dominated by the lower frequency (and equivalently lower field) data. In other words, as more of the lower frequency data are excluded from the fit, the variation of $g_{fit}$ on the fitting range decreases.

The presence of defects and/or interface states can significantly enhance the field required to fully saturate the magnetization relative to a perfect bulk specimen. [37–40] In such cases, $M_{eff}$ is a function of the applied magnetic field until the sample reaches full saturation. The inset of Fig. 10 shows a superconducting quantum interference device (SQUID) magnetization curve for a 20 nm $Ni_{80}Fe_{20}$ sample. The linear diamagnetic background of the Si substrate is determined by fitting a line through the data from 2 T to 4 T, which is then subtracted from the rest of the data. Figure 10 shows a detail of the magnetization curve in the immediate vicinity of $M/M_s = 1$ for positive applied magnetic fields. These data show that even when the sample appears to be fully saturated at fields well below 50 mT, a small amount of the moment is not saturated. Even at applied magnetic fields in excess of 1 T, approximately 0.1 % of the moment is not yet saturated.

A slow approach to saturation is commonly observed in amorphous magnetic materials. In fact, various functional forms to the approach to saturation are found in these materials, with $1/\sqrt{H}$, $1/H$ and $1/H^2$ dependencies for varying types of defect.[37–40] A similar behavior was also observed in crystalline thin films of magnetic oxides and intermetallics, where the presence of antiphase boundaries results in a slow approach to saturation for the spins at the boundary.[41,42] However, in these cases, the slow approach to saturation consists of a large fraction of the total moment relative to what we observe in $Ni_{80}Fe_{20}$. In addition, the simple solid-solution, face-centered cubic structure of Permalloy does not admit such antiphase boundaries found in more complex magnetic oxides. However, these works do show that the presence of defects within the structure and at the interface can generate a slow approach to saturation for a substantial fraction of the spin in the magnetic material.

Another contribution to the unsaturated moment is the presence of magnetization ripple.[43] Observations of ripple domains in $Ni_{80}Fe_{20}$ at fields in excess of the nominal saturation field—as determine from magnetometry—have been previously observed in $Ni_{80}Fe_{20}$ by use of Lorentz microscopy.[44-47] The ripple domain state results from spatial variations of magnetic properties such as crystalline anisotropy, even in materials such as permalloy that exhibit very weak uniaxial anisotropy.[48,49] Since the magnitude and structure of magnetization ripple is a function of film thickness, this effect is consistent with the thickness dependence of the asymptotic behavior of the $g$–factor shown in Fig. 4.

To determine if extrinsic effects can also influence the asymptotic dependence of $g_{fit}$ on $f_{up}$, we investigated whether the interface has an influence over the asymptotic behavior. We fabricated a series of 6 nm thick $Ni_{80}Fe_{20}$ layers with different capping layer materials. In addition, we fabricated a $Ni_{80}Fe_{20}$ layer without the Ta seed layer. Figure 11 shows a chart of the slope taken from the $g_{fit}$ versus $1/f_{up}^2$ curves for the different capping and seed layer conditions. There is a substantial dependence of both the magnitude and sign of the slope on the seed- and cap-layer materials. The largest change occurs for the Pd capping layer. It is well known that Pd polarizes at the interface to ferromagnetic metals.[50-55] In addition, the ferromagnetic transition metal/Pd interface can exhibit significant interface anisotropy.[52,54,56-60] As such, it is possible that the spins at the Permalloy/Pd interface exhibit increased local pinning and/or anisotropy.

Much of the phenomena we observe can also be explained by the presence of a rotatable anisotropy where the easy axis can change direction with the application of a magnetic field and its history. A rotatable anisotropy has been observed in Permalloy thin films.[61] In fact, the magnitude of the rotatable anisotropy was found to be a function of the Permalloy thickness,[62] which is consistent with the thickness variation seen in Fig. 4, as well as the presence of the asymptotic behavior in the out-of-plane geometry. Rotatable anisotropies are also commonly observed in exchange bias multilayer systems or in systems where spins in a thin magnetic layer experience strong pinning at a surface or interface.[63] This is also consistent with the variation of the slope we observe for different seed and capping layers (Fig. 11).

Determination of the exact Kittel equation needed for a given sample is complicated by the existence of multiple in-plane- and out-of-plane-anisotropy terms of varying symmetry for a variety of materials, the possible presence of rotatable anisotropy, and a slow approach to full saturation due to pinning at defects, grain boundaries, and interfaces. In addition, as more fitting parameters are included in the Kittel equation, the unique determination of the parameters by measurements over a limited range of frequencies is challenging. However, we have demonstrated that an accurate value of the $g$–factor can still be obtained by the described asymptotic method, even when the exact form of the anisotropy energy is ambiguous, making this method a particularly powerful means of determining the ratio of orbital to spin moment in polycrystalline alloys of interest for spintronic applications.

**Summary**

In summary, we demonstrated that the fitted value of the spectroscopic $g$–factor asymptotically converges to a fixed value with increasing fitted frequency range. By use of a broadband FMR technique, this asymptotic value of the $g$–factor can be determined with an accuracy approaching 1 part in $10^3$. For sputtered Permalloy films of various thicknesses, $g_{fit}$ converges in proportion to the inverse square of the upper value of the frequency used in the fit. We find a variation of $g$ with the thickness of the Permalloy as well as the measurement geometry, which is consistent with the asymmetry of the orbital moment at interfaces. We supported our assumption that the measured asymptotic value of $g$ is accurate by application of our analysis to simulated data wherein controlled degrees of systematic error and noise could be introduced, and their impact on the fitting procedure could be unambiguously determined. In particular, we find that our asymptotic approach is robust against uncertainties in the functional form of the anisotropy fields that establish the ground state energy of the measured spin system. In addition to providing a method to determine the $g$–factor, the findings of this work suggest that the substantial variation of reported values of the spectroscopic $g$–factor in the literature could in part stem from the diversity of measurement parameters and fitting methods used in the past that have not fully accounted for systematic errors that can be introduced when the anisotropy energy landscape is not fully characterized. While a full determination of the functional form of the anisotropy for our samples was beyond the scope of this study, we show evidence for both rotatable anisotropy and slow saturation, as well as extrinsic dependencies on the choice of seed- and capping-layers .

*Contribution of the National Institute of Standards and Technology, not subject to copyright.

**Figure Captions**

**Figure 1.** Resonance field versus frequency taken of the 10 nm thick $Ni_{80}Fe_{20}$ sample for the in-plane geometry. Examples of the real and imaginary spectra for 8 GHz and 55 GHz frequencies are included in the plot.

**Figure 2.** Plot of the fitted value of the $g$–factor of the 50nm NiFe sample as a function of the upper value of the frequency $f_{up}$ used in the fit to Eq. (2). The insets show $H_{res}$ vs. $f$ and respective fits to Eq. (2) for the data range used to obtain the solid symbols. The horizontal line indicates the asymptotic value of the $g$–factor that the data approaches as $f_{up} \to \infty$.

**Figure 3**. (a) Log-log plot of the different between $g_{fit}$ and $g$ versus $f_{up}$. Plots of the (b) fitted value of the in-plane $g$–factor and (c) fitted value of $M_{eff}$ as a function of the inverse square of the upper value of the frequency used in the fit for the 50 nm $Ni_{80}Fe_{20}$ sample. Linear fits to the data are included as the solid lines.

**Figure 4**. Plots of the fitted value of the in-plane $g$–factor as a function of $1/f_{up}^2$ obtained for the (a) 5nm, (b) 10 nm, (c) 20 nm, and (d) 50 nm $Ni_{80}Fe_{20}$ samples. Data and fits for three values of $f_{low}$ are included in each plot.

**Figure 5.** Plots of (a) $g$ and (b) $M_{eff}$ as a function of the reciprocal thickness of the $Ni_{80}Fe_{20}$ layer for both the in-plane and out-of-plane geometries. Linear regression fits are included as the lines through the respective data.

**Figure 6.** The fitted value of the g-factor when $H_k$ is included as a fitting parameter versus (a) $f_{up}$, and (b) $1/f_{up}^{1.2}$. (c) Plot of the fitted value of $\mu_0 H_k$ versus $f_{up}$. (d) In-plane magnetization curves for the 50 nm sample along the easy and hard magnetic axes.

**Figure 7**. (a) The resonance field as a function of frequency for the out-of-plane geometry along with the fit to the Kittel equation for the 20 nm thick sample. Plots of the fitted value of the $g$–factor in the out-of-plane geometry as a function of $1/f_{up}^2$ used in the fit for the (a) 10 nm and (b) 20 nm $Ni_{80}Fe_{20}$ samples. Data and fits for four values of $f_{low}$ are included in each plot.

**Figure 8**. Plots of $g_{fit}$ as a function of $1/f_{up}$ for the simulated data when $H_k$ is excluded from the fit for the case of (a) $N_H = 0$, and (b) $N_H = 1$ mT, and (c) $H_k$ is included in the fit for the case of $N_H = 1$ mT.

**Figure 9**. Plots of the determined values of g for the simulated data as a function of $H_k$ for the case of $H_k$ being excluded from the fitting function for several values of $N_H$.

**Figure 10.** $M/M_s$ in the vicinity of full magnetic saturation. The inset shows the complete SQUID magnetization curve versus applied magnetic field for the 20 nm $Ni_{80}Fe_{20}$ sample.

**Figure 11.** Chart of the slope taken from the asymptotic $g_{fit}$ versus $1/f_{up}^2$ curves for a fixed value of $f_{low}$ = 4 GHz for different capping/seed layers.

Figure 1

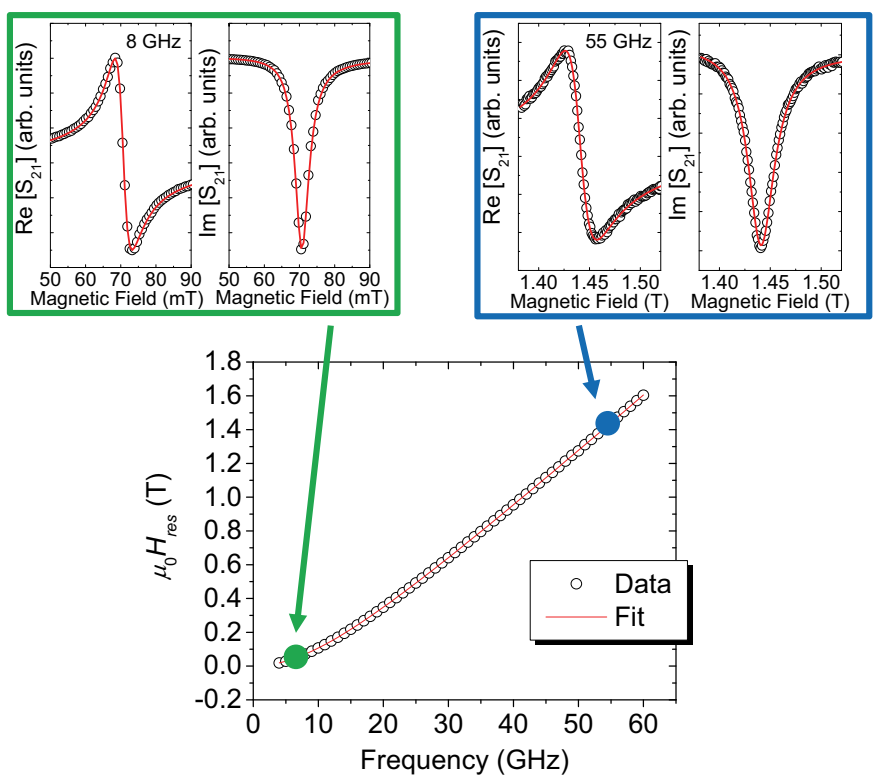

Figure 2

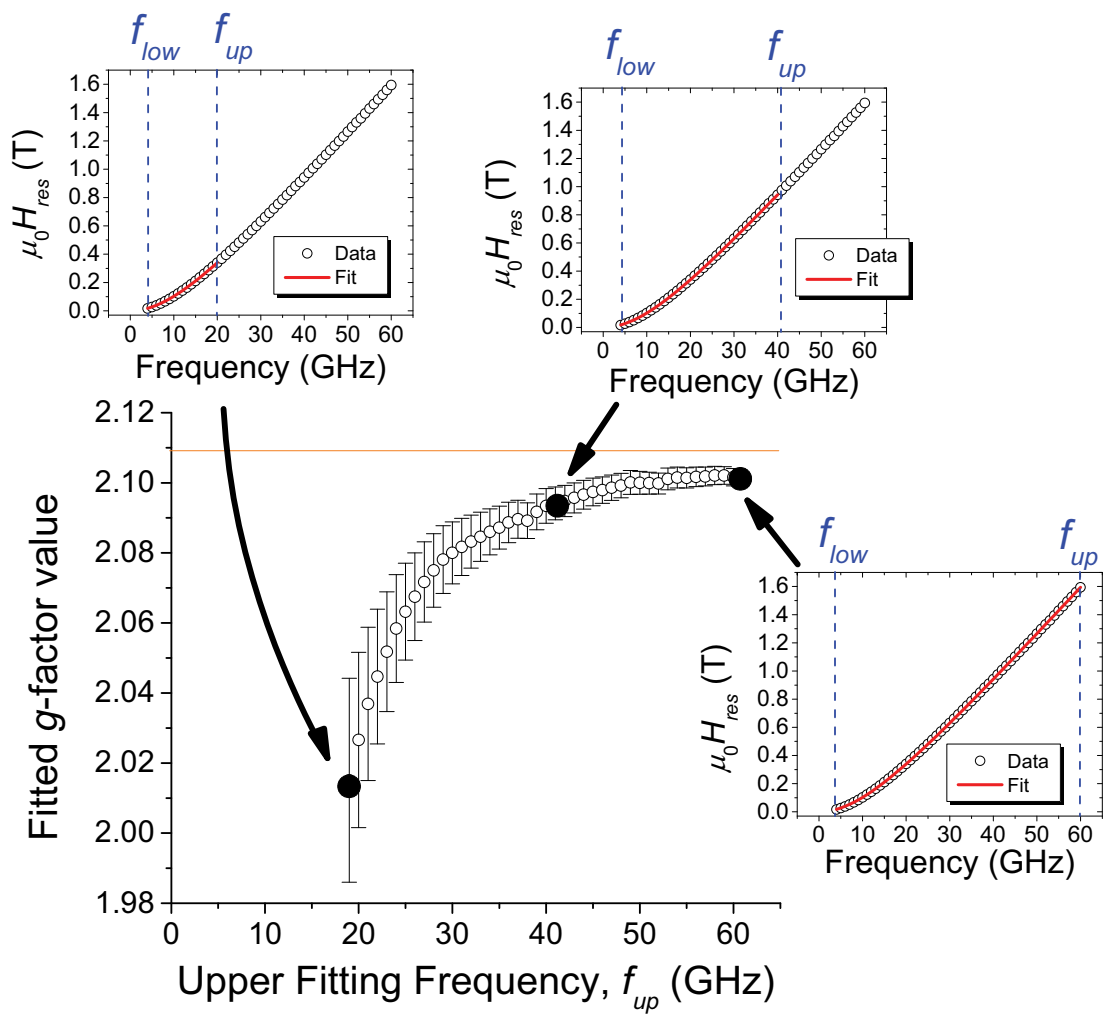

Figure 3

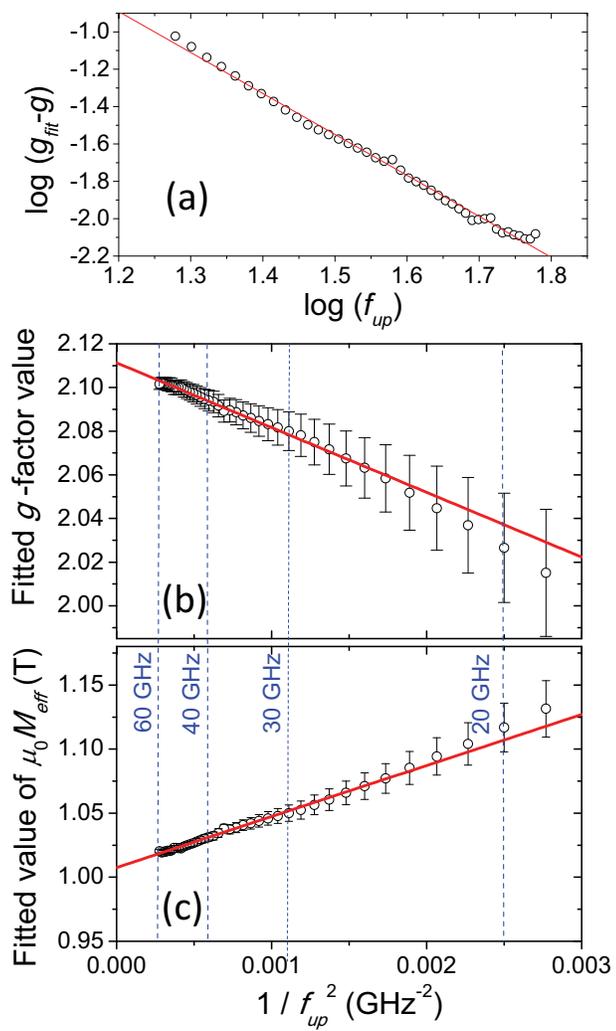

Figure 4

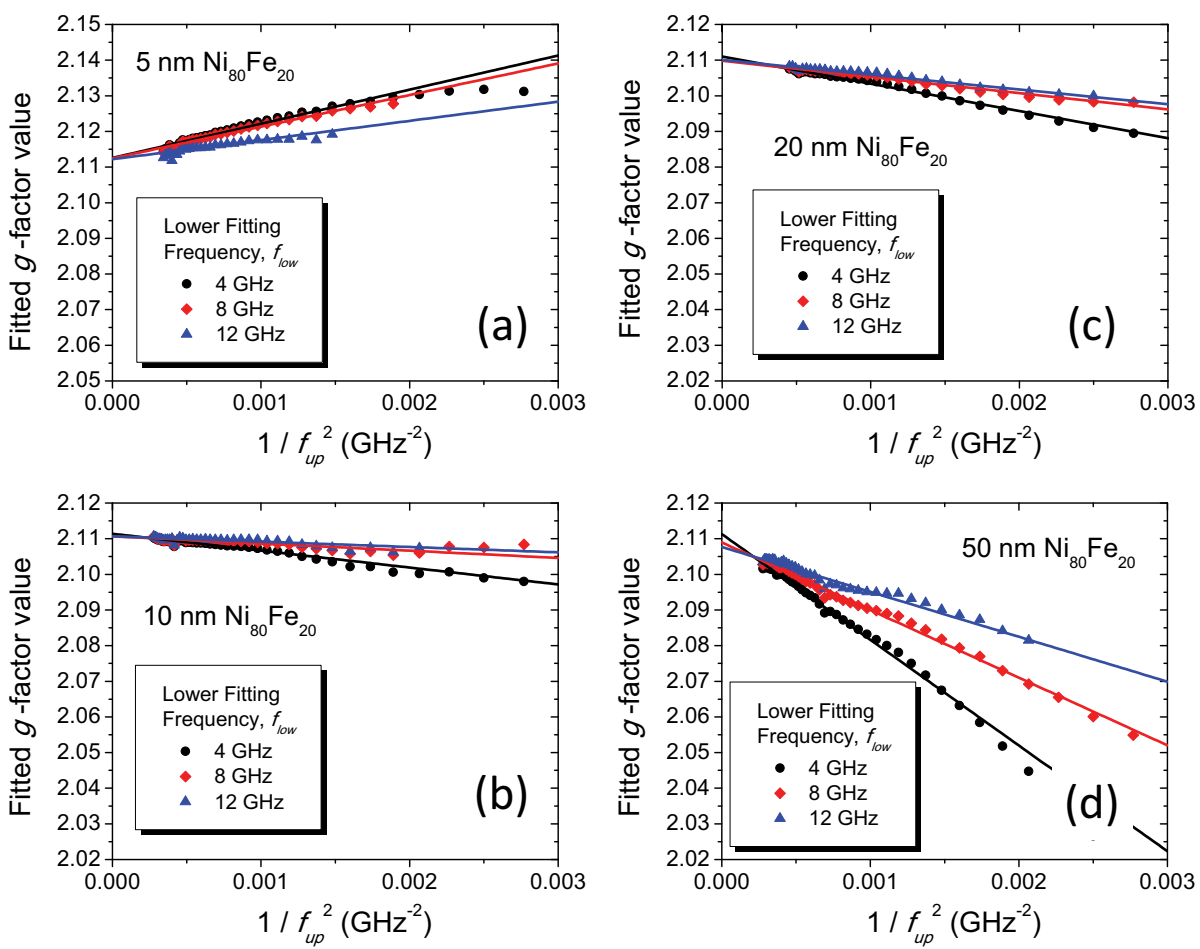

Figure 5

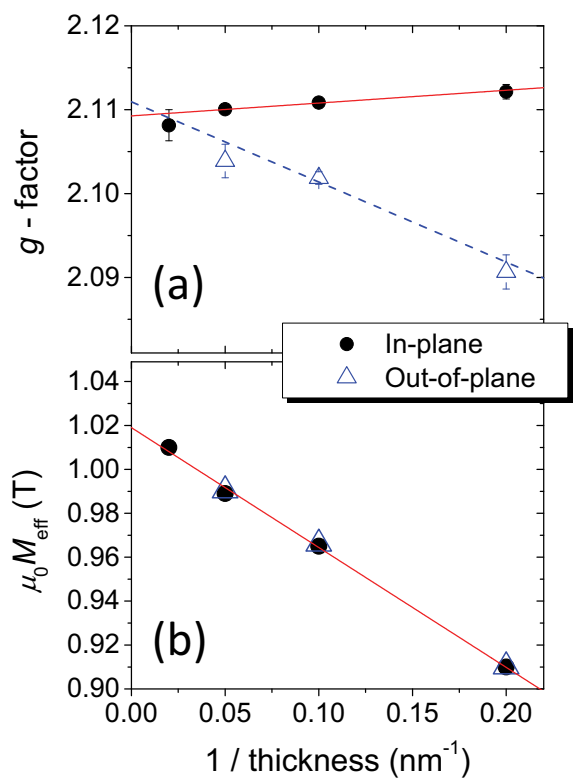

Figure 6

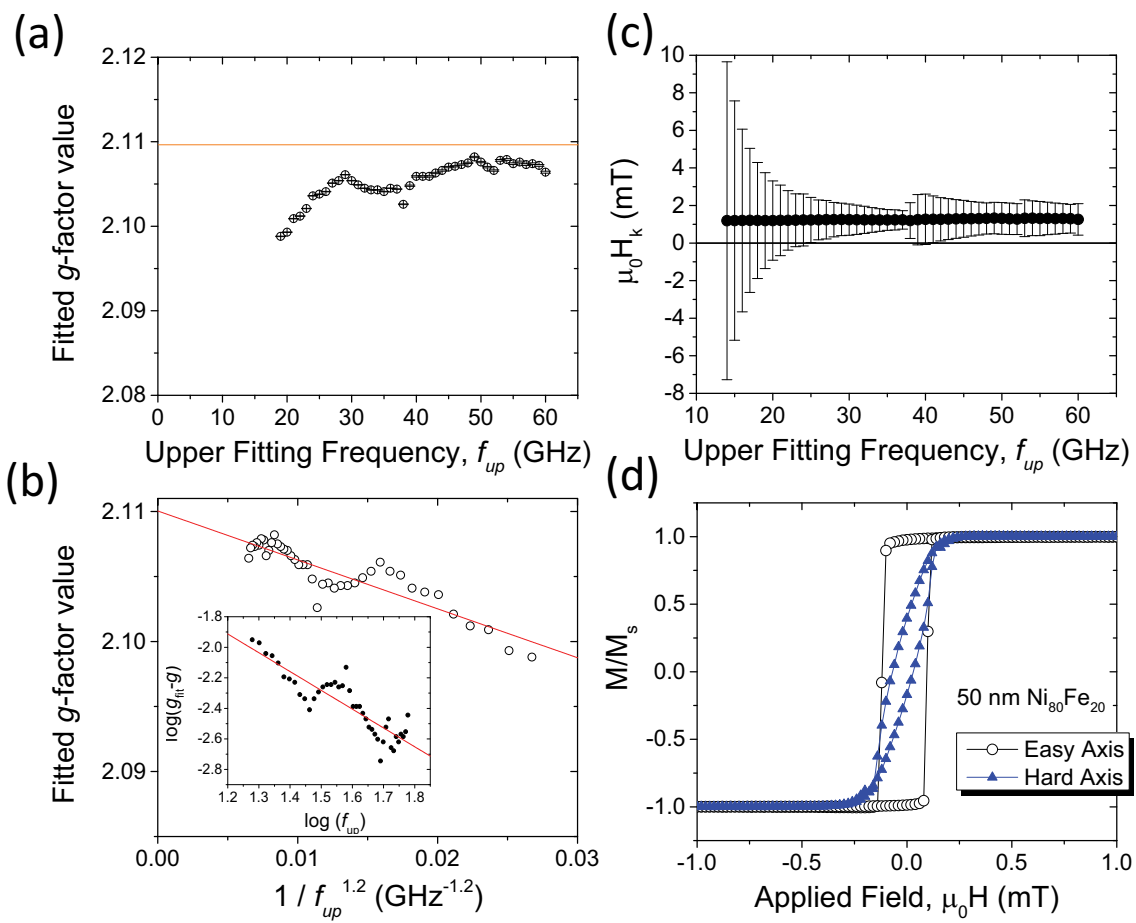

Figure 7

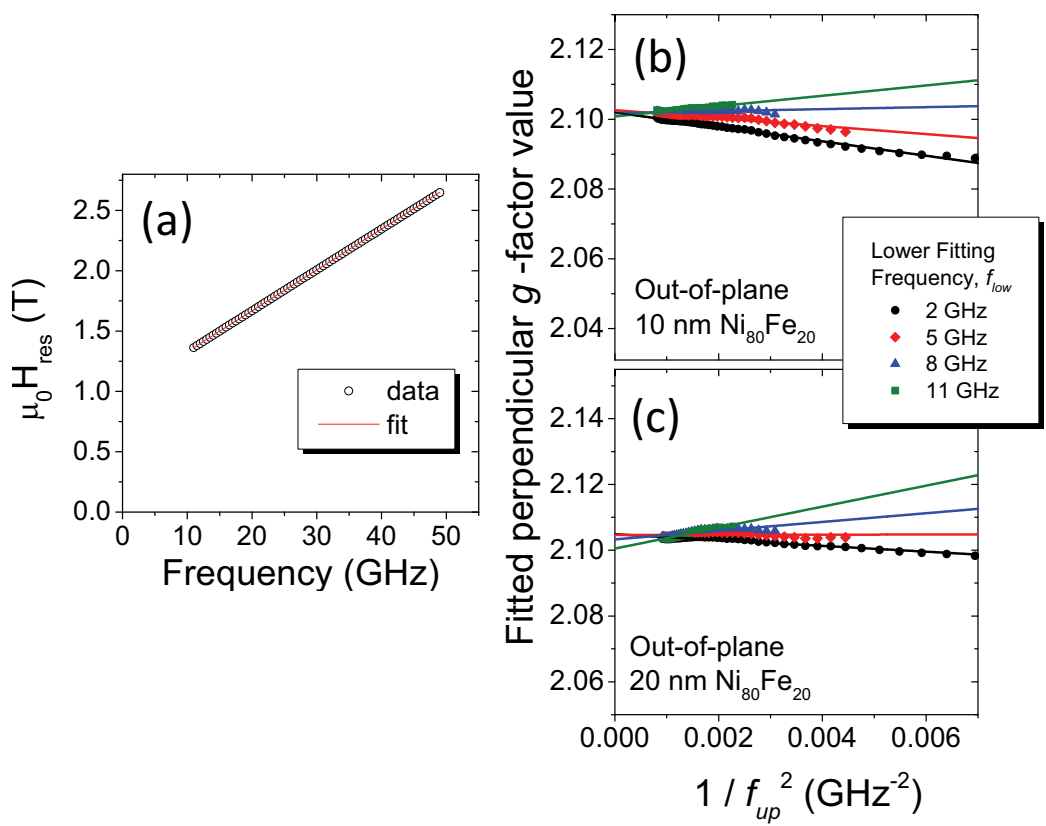

Figure 8

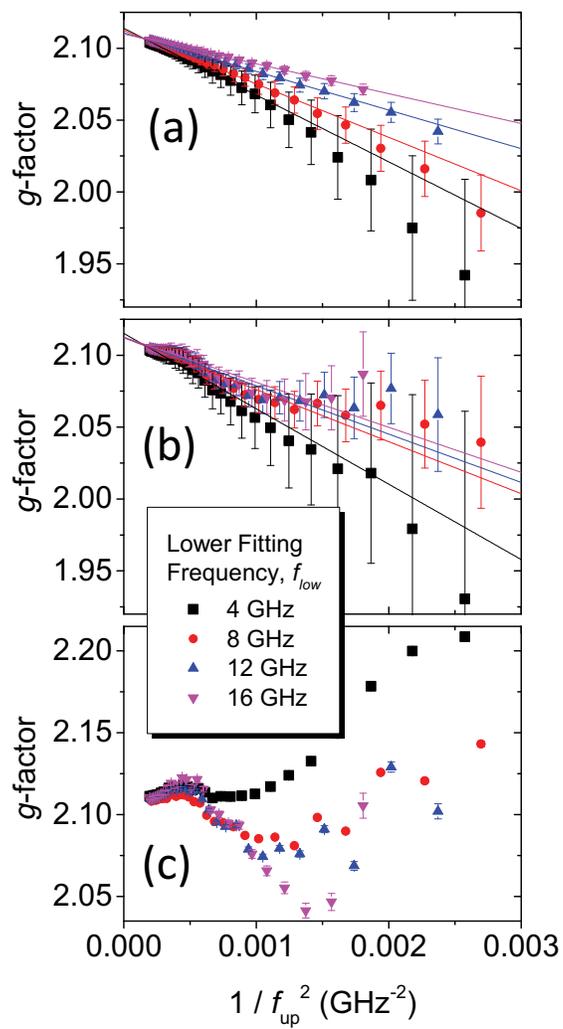

Figure 9

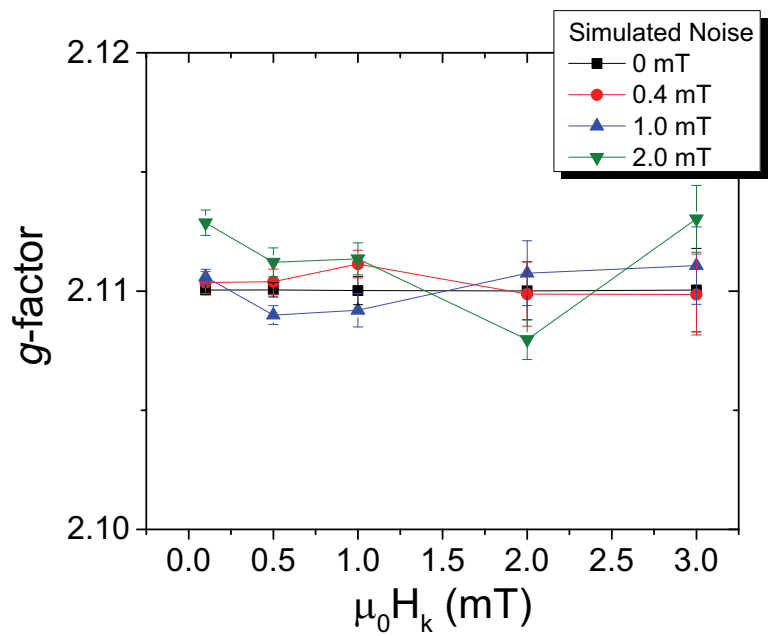

Figure 10

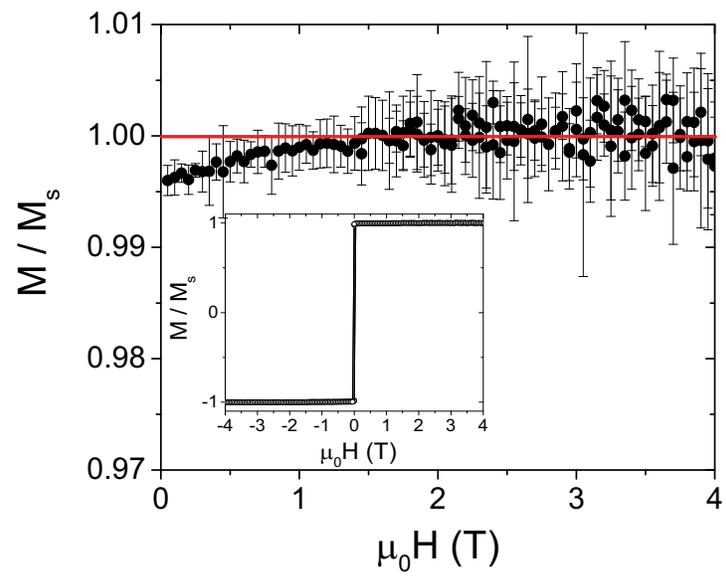

Figure 11

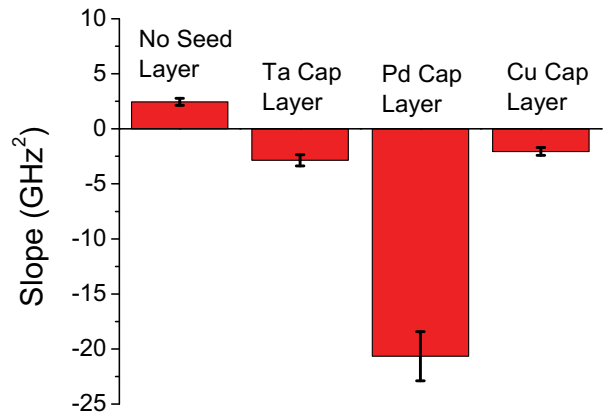